\documentclass[%
 reprint,
 amsmath,amssymb,
 aps,
prb,
showpacs,
longbibliography
]{revtex4-2}

\usepackage{graphicx}
\usepackage{dcolumn}
\usepackage{bm}
\usepackage{physics}
\usepackage{hyperref}
\hypersetup{
    colorlinks,
    linkcolor={blue},
    urlcolor={blue},
    citecolor={blue},
}
\usepackage{color}

\begin{document}

\preprint{APS/123-QED}

\title{Anomalous thermal broadening from an infrared catastrophe in \texorpdfstring{\\}{} two-dimensional quantum antiferromagnets}

\author{Matthew C. O'Brien}
\email{matthew.obrien@unsw.edu.au}
\author{Oleg P. Sushkov}
\affiliation{%
School of Physics, The University of New South Wales, Sydney, NSW 2052, Australia
}%

\pacs{75.10.Jm, 75.30.Ds, 75.30.Gb, 75.50.Ee}

\begin{abstract}
The nature of quasiparticles in 2D quantum antiferromagnets at finite temperature remains an open question despite decades of theoretical work. In particular, it is not fully understood how long wavelength excitations contribute to significant broadening of the experimentally observable spectrum. Motivated by this problem, we consider the $XY$ model of easy-plane antiferromagnets, and compute the dynamic structure factor by direct summation of diagrams. In doing so, we find that non-interacting quasiparticles with infinite lifetimes can still lead to a broad response. This forms the basis for a new paradigm describing the interaction of experimental probes with a physical system, where broadening is due neither to the lifetime, nor to the emergence of fractional quasiparticles. Instead, strong fluctuations drive the probe to absorb and radiate an infinite number of arbitrarily low energy quasiparticles, leading us to draw parallels with the infrared catastrophe in quantum electrodynamics.
\end{abstract}

\maketitle 

\section{\label{sec:intro} Introduction}

The elementary excitations (quasiparticles) in a physical system are always studied experimentally with an external probe, for example, using inelastic neutron or electron scattering, or angle-resolved photoemission spectroscopy. The most common theoretical paradigm (1) for interpreting the results of such an experiment is that the probe measures the imaginary part of the quasiparticle Green's function, which reveals the particle dispersion and lifetime. Recently, a new paradigm (2) has emerged following the discovery of fractional quasiparticles in systems with strong electron correlations. In that case, the bare particle consists of several quasiparticles, and hence, the system's response to an external probe does not have a pole. Well known examples of fractional quasiparticles include spinons in the half-integer spin Heisenberg chain \cite{Faddeev1981}, and spin-charge separation in the 1D non-Fermi (Luttinger) liquid \cite{Schmidt2010}. Critical deconfinement of spinons also belongs to this paradigm \cite{Senthil2004}. However, a subtlety exists in the case of the Luttinger liquid as the behavior of the spectrum depends on the physical nature of the probe. While photoemission couples directly to the bare particle --- placing it in the second paradigm --- an experiment which couples only to the charge degree of freedom falls into the first category since it interrogates the elementary fractional quasiparticles.

In this work, we demonstrate the existence of a new paradigm (3) where the nature of quasiparticles remains unchanged (no fractionalization), and no local measurement can couple directly to the elementary quasiparticles, yet a physical probe necessarily excites an infinite number of quasiparticles, producing a very broad spectral response. This is fundamentally a phenomenon of infrared physics, requiring particles with a gapless dispersion; these are typically gauge or Goldstone bosons. In fact, this behavior has been known in the case of gauge bosons for some time. The infrared catastrophe in quantum electrodynamics posed the problem that radiative corrections (\textit{bremsstrahlung}) to electron scattering cross sections are infrared divergent. Bloch and Nordsieck showed that elastic scattering of electrons is impossible, that the probability of radiating any finite number of photons was zero, and that in fact, electrons must always emit an infinite number of arbitrarily low energy photons \cite{Bloch1937}. This phenomenon occurs for any massless gauge particle, for example, the graviton \cite{Weinberg1965}, but not for particles such as the $W^\pm$ or $Z$ bosons, which acquire mass below the electroweak phase transition. In QED, the effect is weak due to the smallness of the fine structure constant $\alpha \approx 1/137$, and is typically only significant at high energies \cite{Aubert1974,Augustin1974,Greco1975}. In this work, we demonstrate for the first time the possibility of Bloch-Nordsieck-like physics with Goldstone quasiparticles. 

A 2D quantum antiferromagnet (2DQA) --- an easy-plane $XY$ system with in-plane $O(2)$ rotational symmetry, or the fully rotationally symmetric $O(3)$ Heisenberg model --- has long range N\'eel order and gapless Goldstone excitations (magnons) at zero temperature. However, at finite temperature, long range thermal fluctuations destroy the ordered ground state expected from a mean field analysis. This is a consequence of the Mermin-Wagner theorem at $T \neq 0$ \cite{Mermin1966}, which predicts the absence of long range order in any 2D system with a spontaneously broken continuous symmetry. This is why, despite decades of research \cite{Chakravarty1989,Tyc1989,Ty1990,Chubukov1994} (see also Ref. \cite{Sachdev2011} for a review), the physics of 2DQAs remains not fully understood. There is no possibility of fractionalization in the infrared sector --- as opposed to higher energy regions of the magnon dispersion \cite{DallaPiazza2015,Shao2017} --- so the previous analysis of 2DQAs was based on the first paradigm illustrated above; the width of the structure factor was assumed to be equal to the inverse lifetime of magnons. An inconsistency was observed but remains unexplained in the $O(3)$ Heisenberg antiferromagnet: Analytical calculations of the inverse lifetime predicted a small value, while numerics showed a broad structure factor \cite{Ty1990}.

To demonstrate paradigm (3) in 2D quantum antiferromagnets, we consider the exactly integrable $XY$ $\sigma$ model. In this system, the quasiparticles are non-interacting, and hence, have an infinite lifetime. Therefore, the imaginary part of the quasiparticle Green's function has zero width, even at finite temperature. Nevertheless, the spin structure factor at $T \neq 0$ is broad. By direct summation of diagrams, we demonstrate that this broadening is caused by the absorption and stimulated emission of an infinite number of quasiparticles. We call this effect thermal \textit{bremsstrahlung}. While previous analyses typically used an imaginary time path integral formalism, our work uses a modern approach to thermal field theory in real time, which emphasizes the nonequilibrium kinetics of quasiparticles \cite{ Scammell2017}.

This paper is structured as follows: In Section \ref{sec:XY}, we introduce the field-theoretical model and techniques used in the subsequent discussion and present a brief analysis of the zero temperature dynamics to better illustrate the effects of temperature. In Section \ref{sec:temperature}, we obtain an exact diagrammatic expansion of the unpolarized dynamic spin structure factor at finite temperature and carefully separate diagrams with equivalent degrees of divergence in order to resum the structure factor to all orders. We also validate our calculation by verifying the spectral sum rules and comment on the physical implications of the diagrammatic expansion. Section \ref{sec:conclusion} presents our conclusions.



\section{\label{sec:XY} The \texorpdfstring{$XY$}{XY} Model \& Quasiparticles at Zero Temperature}

The low energy sector of a square lattice easy-plane antiferromagnet is described by the $XY$ Hamiltonian
\begin{equation}
    H = J \sum_{\langle i,j \rangle} S_i^{(x)}S_j^{(x)} + S_i^{(y)}S_j^{(y)}, \label{eq:xyhamiltonian}
\end{equation}
where $\mathbf{S}_i = (S_i^{(x)}, S_i^{(y)}, S_i^{(z)})$ is the spin operator at site $i$, the summation is over pairs of nearest neighbor sites, and $J > 0$ is the exchange coupling constant. At zero temperature, the ground state of the system spontaneously breaks the $O(2)$ rotational symmetry of the Hamiltonian, and the low lying excitations are spin waves, or magnons. Goldstone's theorem guarantees that their dispersion will be gapless \cite{Goldstone1962}. We note that since the $\hat{\mathbf{z}}$ component of the spins do not appear in \eqref{eq:xyhamiltonian}, neither the total nor the projection of spin are good quantum numbers. As such, magnon states are determined entirely by their momentum $\ket{\mathbf{k}}$. It is well known that the long wavelength physics of square lattice 2DQAs is captured by the nonlinear $\sigma$ model (NLSM) \cite{Chakravarty1989,Ioffe1988}. In particular, we are interested in the $O(2)$ NLSM
\begin{equation}
    \mathcal{L} = \frac{1}{2} \rho (\partial_\mu \Vec{n})^2,\qquad\qquad \Vec{n}^2 = 1,
\end{equation}
where $\rho \approx J S^2$ is the spin stiffness constant, $S$ is the total spin per lattice site, $\Vec{n} = (n_x, n_y)$ is the re-scaled order parameter, $\partial_\mu = (c^{-1} \partial_t, \partial_x, \partial_y)$ is the 3-gradient, and $c \approx 2 J S$ is the speed of magnons, which have a linear dispersion $\omega_{\mathbf{k}} = c k$; from here on, we set $c = 1$. It is well known that the dispersion of magnons in the $O(3)$ symmetric ferromagnetic ($J < 0$) Heisenberg model is quadratic. However, it is interesting to note that --- as demonstrated in Appendix \ref{app:ferro} --- magnons in an easy-plane ferromagnet have a \textit{linear} dispersion, so the following results should also hold for this case after taking $\rho \rightarrow \lvert \rho \rvert$. The unit vector constraint can be expressed naturally in polar coordinates $\Vec{n} = (\cos\varphi, \sin\varphi)$, so that
\begin{equation}
    \mathcal{L} = \frac{1}{2} \rho (\partial_\mu \varphi)^2.
\end{equation}
It is important to note that this expression has an additional degree of freedom from requiring $\Vec{n}$ to be a single-valued function of the angle. This leads to topological winding effects which we will comment on in the next section. Importantly, the quasiparticle excitations of the $\varphi$ field are non-interacting.

The experimental observable is the dynamic structure factor
\begin{equation}
    S_{ij}(\mathbf{k},\omega) = \int \dd t \int \dd^2 \mathbf{r}\, \langle n_i(\mathbf{r},t) n_j(0) \rangle\, e^{i(\omega t - \mathbf{k}\cdot \mathbf{r})}, \label{eq:skw}
\end{equation}
where at zero temperature $\langle\, \cdot \, \rangle $ is an expectation value with respect to the ground state $\ket{0}$. For simplicity, we will consider the total response $S(\mathbf{k},\omega) = S_{xx}(\mathbf{k},\omega) + S_{yy}(\mathbf{k},\omega)$ throughout this paper. This would, for example, correspond to a measurement using an unpolarized neutron source in an inelastic scattering spectroscopy experiment. To calculate this expression, we leverage the fact that the system can be diagonalized exactly in terms of the $\varphi$ excitations. The general spectral representation is \cite{Lifshitz1995},
\begin{align}
    S(\mathbf{k},\omega) &= \sum_{\alpha} \lvert \bra{\alpha}\Vec{n}(0) \ket{0} \rvert^2 \nonumber \\ &\qquad\quad \times (2\pi)^3 \delta(\omega - \omega_\alpha) \delta^{(2)}(\mathbf{k} - \mathbf{k}_\alpha),
\end{align}
where $\ket{\alpha}$ is an excited $\varphi$ Fock state, and $\omega_\alpha$ and $\mathbf{k}_\alpha$ are the energy and momentum of that state. Without loss of generality, we choose the direction of spontaneous symmetry breaking to be the $\hat{\mathbf{x}}$ axis. According to our definition of $\Vec{n}$, this corresponds to $\varphi = 0$. Then, by expanding around the vacuum, and if $\ket{\alpha} = \ket{\mathbf{k}_1,\dots,\mathbf{k}_m}$ is an $m$-particle state, it is a simple exercise to verify that
\begin{equation}
    \bra{\alpha} n_i(0) \ket{0} = \pm e^{-\langle\varphi^2 \rangle_0/2} \prod_{j=1}^m \frac{1}{\sqrt{2\omega_{\mathbf{k}_j} \rho}}, \label{eq:zero_temp_matrix}
\end{equation}
if the number of particles is even and $i = x$ or the number of particles is odd and $i = y$, and zero otherwise, and where the overall (irrelevant) sign depends on the number of particles. This is simply a product of wavefunction normalizations --- including the factor of $\rho$ since $\varphi$ is dimensionless --- renormalized by the quantum (zero temperature) fluctuations of the $\varphi$ field
\begin{equation}
    \langle\varphi^2 \rangle_0 = \int_0^\Lambda \frac{\dd^2 \mathbf{q}}{(2 \pi)^2} \frac{1}{2\omega_{\mathbf{q}} \rho} = \frac{\Lambda}{4 \pi \rho},
\end{equation}
where $\Lambda \sim \pi/a$ is an ultraviolet cut-off, and $a$ is the lattice spacing in the microscopic model. Since $\langle\varphi^2 \rangle_0$ is directly proportional to the cut-off, it is physically irrelevant, and we absorb it into a redefinition of the $\Vec{n}$ field via renormalization. This amounts to multiplying $\Vec{n}$ by $e^{\langle\varphi^2 \rangle_0/2}$, somewhat like a quasiparticle residue.

\begin{figure}[!t]
    \centering
    \includegraphics[scale=0.85]{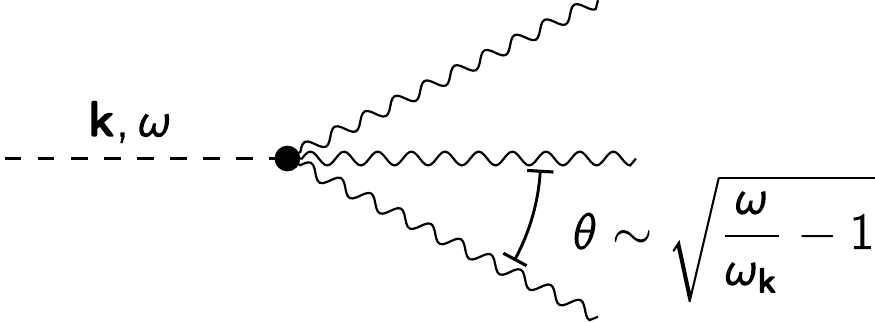}
    \caption{An external source (dashed line) emits three quasiparticles in a narrow cone centered around the direction of momentum transfer $\mathbf{k}$.}
    \label{fig:jet}
\end{figure}

We define the total integrated probability density of exciting $m$ quasiparticles given a transfer of energy $\omega$ and momentum $\mathbf{k}$ to the system from the external source:
\begin{align}
    P_m(\mathbf{k},\omega) &= \frac{1}{m!}\int\dd \lbrace\mathbf{k_\alpha}\rbrace\, \lvert \bra{\mathbf{k}_1,\dots,\mathbf{k}_m}\Vec{n}(0) \ket{0} \rvert^2 \nonumber \\
    &\qquad\qquad \times (2 \pi)^3 \delta(\omega - \omega_\alpha) \delta^{(2)}(\mathbf{k} - \mathbf{k}_\alpha),
\end{align}
where $\dd \lbrace\mathbf{k_\alpha}\rbrace = \prod_{j=1}^m \dd^2 \mathbf{k}_j/(2 \pi)^2$, so that
\begin{equation}
    S(\mathbf{k},\omega) = \sum_{m = 0}^\infty P_m(\mathbf{k},\omega).
\end{equation}
Together, the zero (elastic) and single quasiparticle probabilities yield
\begin{equation}
    S(\mathbf{k},\omega) \propto  \delta(\omega) \delta^{(2)}(\mathbf{k}) + \frac{1}{8 \pi^2 \omega_{\mathbf{k}}\rho} \delta(\omega - \omega_{\mathbf{k}}) ,
\end{equation}
which agrees with the familiar result after restoring units of $c$ \cite{Katanin2011}. However, there is also a multiparticle continuum. The two particle integral can be evaluated exactly:
\begin{equation}
    P_2(\mathbf{k},\omega) = \frac{1}{8 \rho^2 \sqrt{\omega^2 - \omega^2_{\mathbf{k}}}} \, \Theta(\omega - \omega_{\mathbf{k}}), \label{eq:zero_temp_2_particle}
\end{equation}
where $\Theta(x)$ is the Heaviside step function taken with the convention $\Theta(0) = 0$. The three particle integral can also be evaluated exactly:
\begin{equation}
    P_3(\mathbf{k},\omega) = \frac{1}{96 \pi \rho^3}\, \Theta(\omega - \omega_{\mathbf{k}}). \label{eq:zero_temp_3_particle}
\end{equation}
We observe that the probability of each process scales dimensionally as $(\omega/\rho)^{m-3}$. Hence, all higher-order processes are suppressed in the low energy sector $\omega \ll \rho$. Therefore, we find that the zero temperature spectrum consists of the usual $\delta$ peak, as well as a two particle power law tail and a faint multiparticle substrate. We also contrast the direction of quasiparticle emission with the behavior at finite temperature. For $\omega = \omega_{\mathbf{k}}$, conservation of energy and momentum force excited particles to all be emitted in the same direction. However, for $0 < \omega - \omega_{\mathbf{k}} \ll \omega_{\mathbf{k}}$, a region of phase space opens up, and the particles are emitted in a very narrow cone centered around the direction of momentum transferred from the source, as shown in Fig. \ref{fig:jet}. In addition, the total probability is dominated by processes in which the energy is evenly distributed among the excited quasiparticles.

The key message of this section is that quasiparticles are perfectly well-defined at zero temperature. All multiparticle contributions to the spectrum originate from the highly nonlinear coupling of the physical source (neutron magnetic moment) to the harmonic excitations of the system. While these observations are not exactly novel, they form an important basis for the language we will use to describe the kinetics of quasiparticles at finite temperature in the following section.


\section{\label{sec:temperature} Finite Temperature Solution}

For completeness and transparency, we begin by commenting on the topological effects first illustrated by Berezinskii \cite{Berezinskii1971}, Kosterlitz and Thouless \cite{Kosterlitz1973}. It is well known that a single vortex --- a net winding of the angle $\varphi$ --- has energy
\begin{equation}
    E_{\mathrm{vortex}} = E_{\mathrm{core}} + \pi \rho \log \frac{L}{a},
\end{equation}
where $L$ is the linear dimension of the system, the lattice spacing $a$ serves as an ultraviolet cut-off, and $E_{\mathrm{core}} \sim \rho$ is the contribution from inside this region. For a small source energy transfer $\omega \ll \rho$, it is clear that vortices will play no role in the zero temperature dynamics. However, statistical contributions must be taken into account at finite temperature. By expressing the entropy as $S = T\log\Omega$, where $\Omega$ is the number of possible configurations, we estimate the free energy per vortex $F = E - TS$ by packing vortex cores of radius $a$ into a region of area $L^2$:
\begin{equation}
    F_{\mathrm{vortex}} \simeq E_{\mathrm{core}} + \pi \rho \log \frac{L}{a} - T \log \frac{L^2}{a^2}.
\end{equation}
This expression indicates a phase transition temperature $T_{\mathrm{BKT}} = \pi \rho/2$, above which the generation of vortices is thermodynamically favorable. However, by working in the low temperature regime $T \ll \rho$, we will always sit below $T_{\mathrm{BKT}}$, and can safely neglect any topological effects; vortices do not exist at thermal equilibrium. This defines the domain of validity of the following analysis.

\begin{figure*}
    \centering
    \includegraphics[scale=0.85]{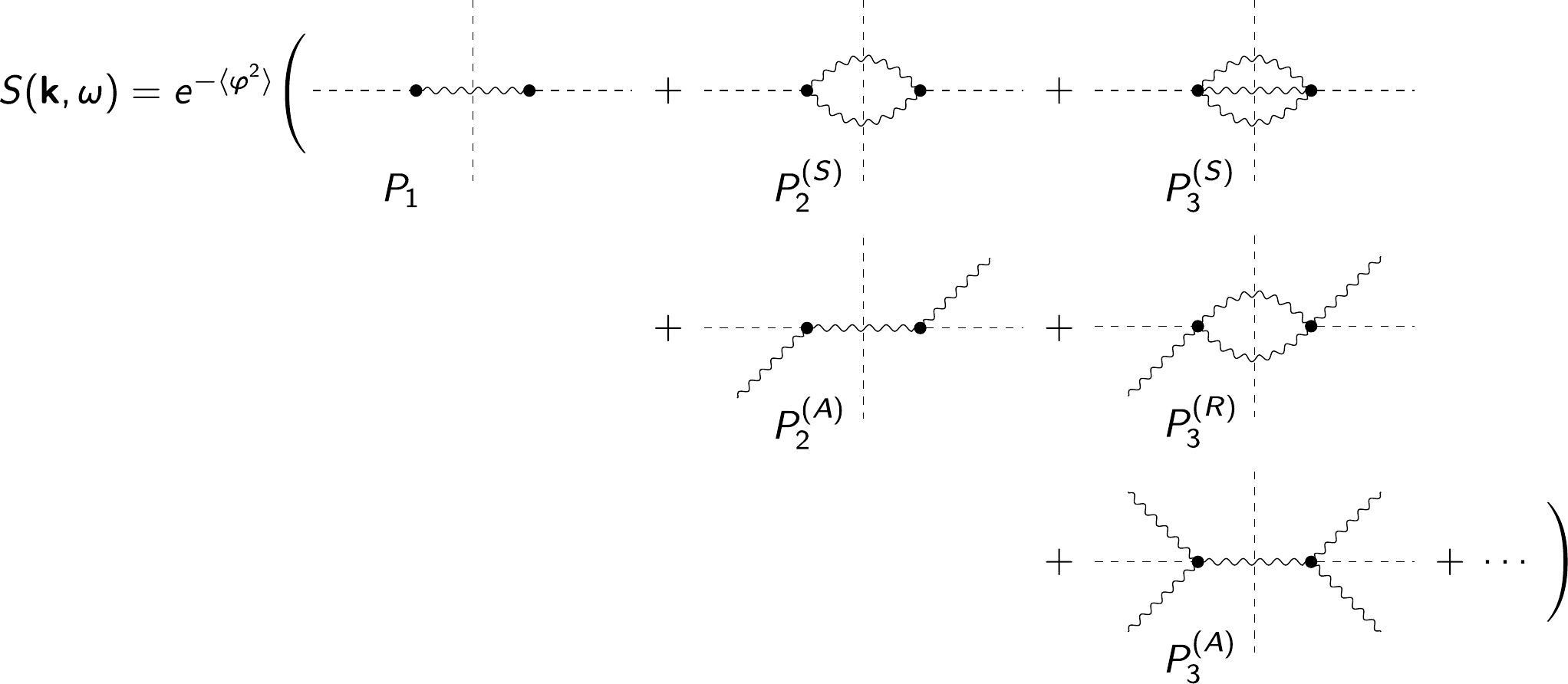}
    \caption{Diagrammatic expansion of the dynamic structure factor. Dashed lines denote the source, wavy lines denote quasiparticles, and vertical dashing denotes application of cut rules. Diagrams are grouped into columns according to the number of quasiparticles in each piece after cutting. $(S)$ denotes Stokes ($\omega > \omega_{\mathbf{k}}$ only), $(A)$ denotes anti-Stokes ($\omega < \omega_{\mathbf{k}}$ only), and $(R)$ denotes Raman scattering (any $\omega$) processes. The trivial elastic scattering contribution ($\omega = 0$) is not shown.
    }
    \label{fig:diagram_exp}
\end{figure*}

The Mermin-Wagner theorem states rigorously that there cannot exist a phase of spontaneously broken continuous symmetry at any finite temperature in two spatial dimensions \cite{Mermin1966}. This is evident in 2DQAs from the infrared divergence of the thermal fluctuations $\langle \varphi^2 \rangle = \infty$. Unlike the ultraviolet divergence of the quantum fluctuations, this divergence has important physical consequences: We will show that it is impossible for an external source to interact with any finite number of quasiparticles, and that the angular distribution of emitted quasiparticles is uniform in the range $[0,2\pi)$, as opposed to the narrow cone at zero temperature.

To separate the thermal and quantum fluctuations, we first regulate the infrared divergence of $\langle \varphi^2 \rangle$ by giving the $\varphi$ field a mass $\mu$ which is smaller than every other energy scale. Then, assuming that the ultraviolet cut-off $\Lambda \gg T$, the thermal fluctuations are
\begin{equation}
    \langle \varphi^2 \rangle - \langle \varphi^2 \rangle_0 = \int_0^\Lambda \frac{\dd^2 \mathbf{q}}{(2 \pi)^2} \frac{n(\omega_{\mathbf{q}})}{\omega_{\mathbf{q}} \rho} \simeq \frac{T}{2 \pi \rho} \log \frac{T}{\mu}, \label{eq:thermal_fluct}
\end{equation}
where $n(\omega) = (e^{\omega/T} - 1)^{-1}$ is the Bose distribution. Since \eqref{eq:thermal_fluct} is independent of $\Lambda$, the effects of the infrared divergence cannot be accounted for by renormalization. However, we renormalize out the quantum fluctuations, and from hereon denote the thermal fluctuations by $\langle\varphi^2\rangle$. Importantly, \eqref{eq:zero_temp_matrix} suggests that all transition matrix elements between quasiparticle number states vanish in the limit that the fictitious mass $\mu \rightarrow 0$. This situation is reminiscent of the infrared catastrophe in quantum electrodynamics, where scattering matrix elements must be resummed to obtain a finite cross section \cite{Yennie1961,Eriksson1961}. In this section, we will show that infrared divergent thermal fluctuations lead to the same physics.

\subsection{Diagrammatic expansion of structure factor}

The dynamic structure factor at finite temperature is still given by \eqref{eq:skw}, though with averaging $\langle\,\cdot\,\rangle$ taken with respect to Gibbs' distribution. Therefore, the general spectral representation at finite temperature becomes
\begin{align}
    S(\mathbf{k},\omega) = \sum_{\alpha,\beta} & \frac{e^{-\omega_\beta/T}}{\mathcal{Z}}\lvert \bra{\alpha}\Vec{n}(0) \ket{\beta} \rvert^2 \nonumber \\ & \times (2\pi)^3 \delta(\omega - \omega_{\alpha\beta}) \delta^{(2)}(\mathbf{k} - \mathbf{k}_{\alpha\beta}), \label{eq:skw_finiteT}
\end{align}
where $\mathcal{Z}$ is the partition function, $\omega_{\alpha\beta} = \omega_\alpha - \omega_\beta$, and similarly for $\mathbf{k}_{\alpha\beta}$. Performing the Gibbs averaging yields an expansion in terms of effective matrix elements connecting thermal equilibrium ($\varepsilon$) and out-of-equilibrium states ($\alpha$) containing more or fewer excitations than the average occupation:
\begin{align}
    S(\mathbf{k},\omega) &= \sum_{\alpha} \lvert \mathcal{M}(\varepsilon \rightarrow \alpha)\rvert^2 \nonumber \\ &\qquad\quad \times (2\pi)^3 \delta(\omega - \omega_{\alpha\varepsilon}) \delta^{(2)}(\mathbf{k} - \mathbf{k}_{\alpha\varepsilon}), \label{eq:skw_finiteT_2}
\end{align}
where
\begin{equation}
    \lvert \mathcal{M}(\varepsilon \rightarrow \alpha)\rvert^2 = e^{-\langle \varphi^2 \rangle} \prod_{j=1}^m \frac{n(\omega_{\mathbf{k}_j}) + \sigma}{2\omega_{\mathbf{k}_j} \rho}, \label{eq:equi_matrix_element}
\end{equation}
$\sigma = 1$ or $0$ if the quasiparticle is excited by the source or absorbed from the thermal bath, respectively, and $\omega_{\alpha\varepsilon} = \sum \omega_{\mathrm{emitted}} - \sum \omega_{\mathrm{absorbed}}$. Due to the prefactor $e^{-\langle \varphi^2 \rangle}$ in \eqref{eq:equi_matrix_element}, all $T > 0$ processes involving a finite number of quasiparticles are forbidden in the limit $\mu \rightarrow 0$. To obtain a physical answer, we factorize out the ``vertex correction'' $e^{-\langle \varphi^2 \rangle}$, and perform a diagrammatic expansion of the ``bare'' terms of the structure factor, as shown in Fig. \ref{fig:diagram_exp}. However, the subtlety lies in the fact that infinitely-many of the bare diagrams are also infrared divergent. In this section, we will demonstrate the solution to this problem by carefully resumming the diagrams to obtain a result which is independent of the infrared cut-off $\mu$.

While the first row of Fig. \ref{fig:diagram_exp} is standard, using cut rules to represent the retarded Green's function identity for the $\Vec{n}$ field \cite{Lifshitz1995},
\begin{equation}
    -\frac{1}{\pi} \Im G^R(\mathbf{k},\omega) = (1 - e^{-\omega/T}) S(\mathbf{k},\omega),
\end{equation}
we introduce a new representation for the additional processes allowed at finite temperature (shown in the second and third rows): Firstly, we note that \eqref{eq:skw_finiteT} implies directly that
\begin{equation}
    S(\mathbf{k},-\omega) = e^{-\omega/T}S(\mathbf{k},\omega). \label{eq:negative_w}
\end{equation}
Therefore, we can consider $\omega > 0$ in all the following without loss of generality. In that case, the external source must necessarily absorb a quasiparticle from the thermal bath in order to conserve both energy and momentum. These processes are represented by external quasiparticle lines in Fig. \ref{fig:diagram_exp}. It is clear that the momentum of these ``external'' particles is not fixed since every mode in the system has some finite probability of being occupied. However, they are still real on-shell particles. This diagrammatic expansion allows us to investigate the origins of different infrared divergences.



Importantly phase space is independent of temperature. Therefore, since multiparticle processes are forbidden for $\omega = \omega_{\mathbf{k}}$ (as noted in the previous section), the elastic and one particle processes can be treated separately from the others. However, this means that the $\delta$ peaks will be weighted by $e^{-\langle\varphi^2\rangle}$, and hence, destroyed by thermal fluctuations in the limit $\mu \rightarrow 0$; the absence of an elastic scattering peak corresponds exactly to the destruction of the static long range order. The situation is considerably more complicated for $\omega \neq \omega_{\mathbf{k}}$, and so we will focus on the bare multiparticle diagrams and return to the vertex correction once their structure has been made clear.

The first non-trivial processes involve an interaction with two quasiparticles. As at zero temperature, the source can produce two excitations ($P_2^{(S)}$ in Fig. \ref{fig:diagram_exp}). However, a new process is now possible, where one particle is emitted, and a second is absorbed by the source ($P_2^{(A)}$ in Fig. \ref{fig:diagram_exp}). In the high energy limit $\omega \gg T$, the Bose distribution $n(\omega) \simeq e^{-\omega/T}$, and we recover the $T = 0$ results, with exponential suppression of absorption from the thermal bath. In the opposite limit $\omega \ll T$, we expand $n(\omega) \simeq T/\omega$, allowing us to evaluate the probability integrals exactly. Defining the detuning $\Delta = \omega - \omega_{\mathbf{k}}$, we find for the Stokes process
\begin{equation}
    P_2^{(S)}(\mathbf{k},\omega) = \frac{T^2}{2 \rho^2 \Delta(\omega_{\mathbf{k}} + \Delta)(2\omega_{\mathbf{k}} + \Delta)}\, \Theta(\Delta). \label{eq:2_stokes}
\end{equation}
Specifically, the energy of all quasiparticles is bounded below by $\Delta/2$ and above by $\omega_{\mathbf{k}} + \Delta/2$, and the dominant contribution to the integral occurs for one excitation with energy $\sim \omega_{\mathbf{k}}$, and another with energy $\sim \Delta$. Notably, no infrared divergence occurs. This is only due to phase space constraints. Similarly, for the anti-Stokes process, we find
\begin{align}
    P_2^{(A)}(\mathbf{k},\omega) &= \frac{T^2}{2 \rho^2 \lvert\Delta\rvert(\omega_{\mathbf{k}} + \Delta)(2\omega_{\mathbf{k}} + \Delta)}\, \Theta(-\Delta) \nonumber \\
    &\qquad\times \left( \frac{4}{\pi}\arctan\sqrt{\frac{\omega_{\mathbf{k}} + \omega}{\omega_{\mathbf{k}} - \omega}} - 1 \right). \label{eq:2_antistokes}
\end{align}
In this case, the quasiparticle energy is again bounded below by $\lvert \Delta \rvert/2$, though it now has no upper bound since the thermal bath has a finite --- though small --- probability of exciting an arbitrarily high energy particle. Despite this, the dominant contribution still occurs for one particle with energy $\sim \lvert\Delta\rvert$; this reassures us as to the validity of expanding the Bose distributions to leading order in $T/\omega$. Notably, both \eqref{eq:2_stokes} and \eqref{eq:2_antistokes} reduce to the same expression in the limit $\lvert \Delta \rvert \ll \omega_{\mathbf{k}}$, which is the regime we are interested in:
\begin{equation}
    P_2(\mathbf{k},\omega) \simeq \frac{T^2}{4 \rho^2 \omega_{\mathbf{k}}^2 \lvert \Delta \rvert}. \label{eq:2_approx}
\end{equation}
If we started by assuming that one quasiparticle in particular had energy $\ll \omega_{\mathbf{k}}$, then both integrals would yield precisely \eqref{eq:2_approx}. This validates our inference regarding the distribution of energy among the excitations.

The bare expressions for the two particle processes were free of infrared divergences, while the vertex correction remains infinite ($e^{-\langle\varphi^2\rangle} \rightarrow 0$). Therefore, it is necessary to go to the next order to understand the structure of the theory. As shown in Fig. \ref{fig:diagram_exp}, there are three distinct processes involving three particle interactions at finite temperature. Since we have justified the assumption that only one quasiparticle will carry the vast majority of the energy transferred from the source when $\omega \approx \omega_{\mathbf{k}}$, we will proceed by using this approximation. This time however, the available volume of phase space is larger, and we find that the energy of the quasiparticles is only bounded below by the fictitious mass. Specifically, we find that all three processes have the form
\begin{equation}
    P_3(\mathbf{k},\omega) = \frac{T^2}{4 \rho^2 \omega_{\mathbf{k}}^2 \lvert \Delta \rvert} \left( \frac{T}{4 \pi \rho} \log \frac{\lvert \Delta \rvert}{\mu} \right) , \label{eq:finite_T_3_particle}
\end{equation}
though the Stokes process contributes only for $\Delta > 0$, the anti-Stokes process for $\Delta < 0$, while the Raman scattering type process is possible regardless of the detuning. In each case, there will be an emitted particle with energy $\sim \omega_{\mathbf{k}}$, a particle with energy $\sim \lvert\Delta\rvert$, and one with energy in the range $(\mu, \lvert\Delta\rvert)$. Interestingly, \eqref{eq:finite_T_3_particle} has the form
\begin{equation}
    P_2(\mathbf{k},\omega) \times \int_0^{\lvert\Delta\rvert} \frac{\dd^2 \mathbf{q}}{(2 \pi)^2} \frac{n(\omega_{\mathbf{q}}) + \sigma}{2 \omega_{\mathbf{q}} \rho},
\end{equation}
where the difference between emission and absorption (represented by $\sigma = 1 \text{ or } 0$) is irrelevant in the limit $\omega \ll T$. That is, the probability of a three particle interaction factorizes as a two particle interaction along with either the emission or absorption of a very low energy quasiparticle, into or from any direction. Importantly, this additional emission/absorption is statistically independent from the higher energy quasiparticles; near resonance, an additional very low energy particle can take part in the interaction without affecting conservation of energy and momentum. We will return to this point at the end of this section.

In fact, it is clear from the above analysis that any number of additional particles can be involved in the same manner. An $N$ particle interaction (probability $P_N$ in Fig. \ref{fig:diagram_exp}) will contain a single ``hard'' excitation with energy of order $\omega_{\mathbf{k}}$, a second quasiparticle of moderate energy of order $\lvert\Delta\rvert$, and $N - 2$ very ``soft'' quasiparticles. Conservation of energy forces the hard and moderate particles to always be emitted if $\Delta > 0$, and the moderate particle to be absorbed when $\Delta < 0$. Therefore, there will be $N$ distinct bare (without the vertex correction $e^{-\langle\varphi^2\rangle}$) diagrams corresponding to the ways in which the soft quasiparticles can be emitted or absorbed. We must also remember to weight each possible diagram by the correct symmetry factor corresponding to the combinations of identical soft particles; for example, a diagram with $M$ emitted soft excitations should be divided by $1/[(N - 2 - M)! M!]$. From this, we infer that the total (bare) probability for the external source to cause an $N$ particle interaction is
\begin{align}
    P_N^{(\mathrm{tot})} &= P_2(\mathbf{k},\omega) \left( \frac{T}{4 \pi \rho} \log \frac{\lvert \Delta \rvert}{\mu} \right)^{N-2} \nonumber \\
    &\qquad\qquad\times \sum_{M=0}^{N-2} \frac{1}{(N - 2 - M)! M!} \nonumber \\
    &= P_2(\mathbf{k},\omega) \frac{1}{(N - 2)!} \left( \frac{T}{2 \pi \rho} \log \frac{\lvert \Delta \rvert}{\mu} \right)^{N-2}. \label{eq:pn_tot}
\end{align}
Therefore, each additional soft quasiparticle contributes a factor of $\log\mu$. This infrared divergence can only be cured by considering the vertex correction factor. If it is not clear to the reader that the factorization of the probability holds to all orders, we also give a mathematical proof of this identity in Appendix \ref{app:n_particles}.

\subsection{Resummation of diagrams}

\begin{figure}[!ht]
    \centering
    \includegraphics[scale=0.85]{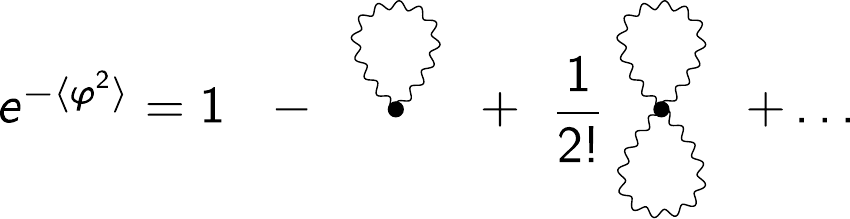}
    \caption{The vertex correction can be expanded as a sum of loop diagrams. Each loop represents a factor of $\langle\varphi^2\rangle$.}
    \label{fig:vertex}
\end{figure}

We are now in a position to investigate the thermal vertex corrections properly. To do so, we must modify the diagrammatic expansion shown in Fig. \ref{fig:diagram_exp}. Firstly, for finite $\mu$, the vertex correction can also be expanded as a sum of diagrams, as shown in Fig. \ref{fig:vertex}.

\begin{figure}[!t]
    \centering
    \includegraphics[scale=0.85]{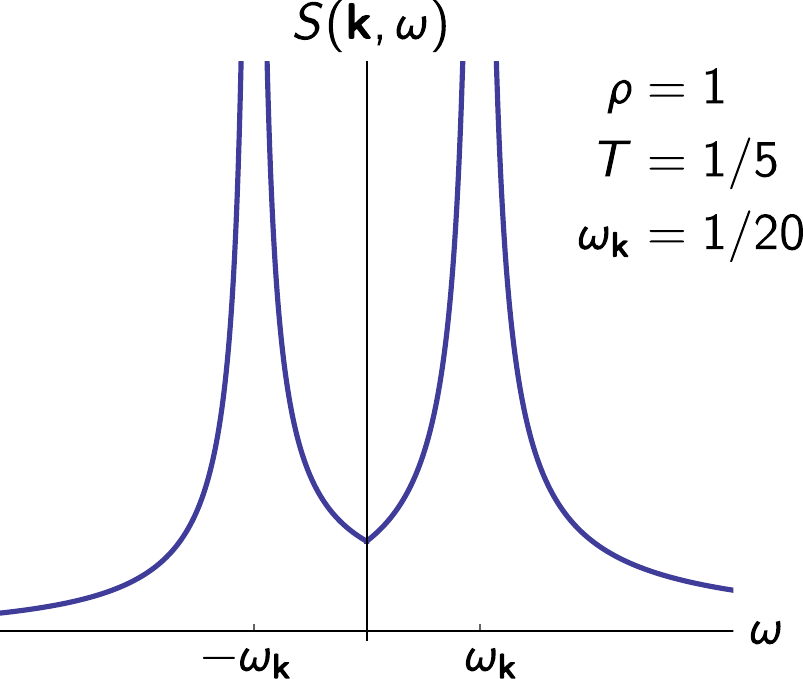}
    \caption{The dynamic structure factor at fixed source momentum transfer $\mathbf{k}$, as a function of energy transfer $\omega$. Continuation of \eqref{eq:skw_summed} to $\omega < 0$ follows from \eqref{eq:negative_w}.}
    \label{fig:spectrum}
\end{figure}

The complete diagrammatic expansion of the dynamic structure factor $S(\mathbf{k},\omega)$ is obtained by inserting the expression from Fig. \ref{fig:vertex} into Fig. \ref{fig:diagram_exp} and distributing terms. Importantly, this approach is not a conventional Feynman expansion, since the coefficients of the loop diagrams in Fig. \ref{fig:vertex} are not equal to unity. Each loop also contributes a factor of $\log\mu$. However, the exponential expansion contains an alternating sign which allows us to regroup the diagrams as follows: A diagram with $L$ loops receives the vertex correction
\begin{equation}
    \frac{(-1)^L}{L!} \langle\varphi^2\rangle^L = \frac{(-1)^L}{L!} \left( \frac{T}{2 \pi \rho} \log \frac{T}{\mu} \right)^{L}. \label{eq:vertex_correct}
\end{equation}
Defining the degree of divergence $D$ of a diagram as the power of $\log \mu$ it contains, we then observe from \eqref{eq:pn_tot} and \eqref{eq:vertex_correct} that $D = N - 2 + L$. Hence, in order to correctly sum the diagrammatic expansion of the dynamic structure factor, we must consider separately each set of diagrams with equal values of $D$. Defining $X = (T/2\pi\rho)\log(\lvert\Delta\rvert/\mu)$, the total order $D$ contribution to the structure factor is

\begin{align}
    S^{(D)}(\mathbf{k},\omega) &= P_2(\mathbf{k},\omega) \sum_{L = 0}^D \frac{(-1)^L}{(D - L)! L!} X^{D - L}\langle\varphi^2\rangle^L \nonumber \\
    &= \frac{1}{D!} P_2(\mathbf{k},\omega) \left( \frac{T}{2 \pi \rho} \log \frac{\lvert \Delta \rvert}{T} \right)^{D}. \label{eq:orderD}
\end{align}
Since \eqref{eq:orderD} is independent of the fictitious mass $\mu$, our result shows that order-by-order, all infrared divergences cancel each other. Physically, this indicates that the interaction of an external source with arbitrarily soft real quasiparticles is indistinguishable from an interaction with the fluctuating thermal bath. Since we have already eliminated the resonant $\delta$ peaks, and having regulated all divergent terms, we sum the diagrams to all orders in $D$ and send $\mu \rightarrow 0$, giving
\begin{equation}
    S(\mathbf{k},\omega) = \frac{T^2}{4 \rho^2 \omega_{\mathbf{k}}^2 \lvert \omega - \omega_{\mathbf{k}} \rvert} \left( \frac{\lvert \omega - \omega_{\mathbf{k}} \rvert}{T} \right)^{T/2\pi\rho}, \label{eq:skw_summed}
\end{equation}
as shown in Fig. \ref{fig:spectrum}. Therefore, the effect of the thermal fluctuations is to renormalize the bare two particle contribution to the spectrum by a power law with a non-universal critical exponent $T/2\pi\rho$. This in itself is also interesting, as $T/\rho$ is the effective coupling constant of the classical statistical mechanical nonlinear $\sigma$ model. This mirrors the situation in quantum electrodynamics, where the distribution of radiated photons reproduces the spectrum of classical \textit{bremsstrahlung} \cite{Bloch1937}.

\subsection{Spectral sum rules}

A structure factor should obey a set of sum rules for the moments of the distribution \cite{Lifshitz1995}. In particular, the dynamic structure factor should reproduce the static factor
\begin{align}
    S(\mathbf{k}) &= \int \frac{\dd \omega}{2 \pi} S(\mathbf{k},\omega)\nonumber \\
    &= \int \dd^2 \mathbf{r}\, \langle \Vec{n}(\mathbf{r},0)\cdot\Vec{n}(0) \rangle\, e^{-i \mathbf{k}\cdot\mathbf{r}}. \label{eq:static}
\end{align}
Therefore, we verify our result \eqref{eq:skw_summed} by calculating $S(\mathbf{k})$ directly from the $\Vec{n}$ field correlation function, using the polar coordinate identity
\begin{equation}
    \langle \Vec{n}(\mathbf{r})\cdot\Vec{n}(0) \rangle = \exp\Big( \langle\varphi(\mathbf{r})\varphi(0) \rangle - \langle\varphi^2\rangle \Big).
\end{equation}
The exponent above is given by
\begin{equation}
    \int_0^\Lambda \frac{\dd^2 \mathbf{q}}{(2\pi)^2} \frac{e^{i\mathbf{q}\cdot\mathbf{r}} - 1}{2 \omega_{\mathbf{q}}\rho} [1 + 2 n(\omega_{\mathbf{q}})].
\end{equation}
The first term in the brackets corresponds to the quantum contributions which we have already renormalized out, and if $r \gg 1/T$, then the complex exponential will oscillate rapidly and average to zero. Therefore, this expression is well approximated in the limit of large $r$ by
\begin{equation}
    -\int_{1/r}^T \frac{\dd^2 \mathbf{q}}{(2\pi)^2} \frac{n(\omega_{\mathbf{q}})}{\omega_{\mathbf{q}}\rho} \simeq - \frac{T}{2 \pi \rho} \log(Tr). \label{eq:phi_corr}
\end{equation}
Hence, we recover the well-known algebraically decaying correlation function \cite{Kadanoff1979},
\begin{equation}
    \langle \Vec{n}(\mathbf{r})\cdot\Vec{n}(0) \rangle \simeq  (Tr)^{-T/2\pi\rho}.
\end{equation}
Therefore, we find that the Fourier transform \eqref{eq:static} is --- by compensating for the sharp $1/r$ cut-off in \eqref{eq:phi_corr} with an exponential regulator,
\begin{equation}
    S(\mathbf{k}) \simeq \frac{T}{\rho k^2} \left( \frac{k}{T} \right)^{T/2\pi\rho}. \label{eq:static_result}
\end{equation}
From \eqref{eq:negative_w}, we see that our result for the dynamic structure factor \eqref{eq:skw_summed} is sharply peaked at $\pm \omega_{\mathbf{k}}$, so we restrict the $\omega$ integral in \eqref{eq:static} to the range $(-2\omega_{\mathbf{k}}, 2\omega_{\mathbf{k}})$, and find exact agreement with \eqref{eq:static_result}.

The total integrated intensity should also be equal to the zero temperature elastic intensity \cite{Katanin2011}, which in this case is normalized to unity. Using $\eqref{eq:static_result}$, which is valid for $k \ll T$, we find that
\begin{equation}
    \int_0^T \frac{\dd^2 \mathbf{k}}{(2 \pi)^2} S(\mathbf{k}) = 1,
\end{equation}
as required. This confirms that the spectral intensity associated with the elastic Bragg peak is transferred entirely to the infinite particle spectrum at finite temperature. In particular, we note that the sum rule would not be satisfied if the radiation of soft quasiparticles was neglected.

\subsection{Physical remarks}

To summarize, we have demonstrated the following key facts:
\begin{enumerate}
    \item At zero temperature, an external probe can scatter elastically from the static antiferromagnetic order and the inelastic spectrum also contains a clear quasiparticle peak.
    \item At finite temperature, elastic scattering and the production of any finite number of quasiparticles is forbidden. In particular, no single quasiparticle spectral weight remains.
    \item The source will both absorb and emit an infinite number of arbitrarily low energy quasiparticles in a statistically independent manner.
\end{enumerate}
We conclude this section by noting that statistical independence indicates the emission (and absorption) of thermal \textit{bremsstrahlung} obeys Poisson statistics. Therefore, the probability that the source will emit exactly $N$ quasiparticles into the energy interval $0 < E_- < E_+ \ll \Delta$ will be given by
\begin{equation}
    P(\text{emit }N\varphi) = \frac{1}{N!} \lambda^N e^{-\lambda},
\end{equation}
where
\begin{equation}
    \lambda = \frac{T}{4 \pi \rho} \log \frac{E_+}{E_-},
\end{equation}
is the mean number of emitted quasiparticles with energy in the range $(E_-,E_+)$. As a consistency check, we see that $P(N)$ vanishes if $E_- = 0$, since no finite number of arbitrarily soft excitations can be emitted. Radiated soft photons in quantum electrodynamics are also Poisson-distributed \cite{Berestetskii1982}.




\section{\label{sec:conclusion} Conclusions}

In this paper, we have demonstrated a new paradigm describing the relationship between an experimental probe and the elementary quasiparticles of a physical system. We emphasize that the example of the $XY$ model discussed is not a manifestation of the fractional quasiparticle paradigm for the following reasons: We have shown that at zero temperature the structure factor has a pole corresponding to free elementary excitations, and that the presence of a multiparticle tail is only due to the nonlinear coupling between the probe and the $\varphi$ field. Additionally, since the quasiparticles are free, their properties are completely unchanged by temperature. Therefore, the spectral width at finite temperature cannot be explained as the original $\Vec{n}$ field fractionalizing into infinitely-many $\varphi$ particles. Instead, this anomalous broadening is driven by long wavelength fluctuations, and specifically, the enhancement of these fluctuations in two dimensions. Just as charged particles are dressed by quantum fluctuations of the electromagnetic field, an external physical probe --- which couples to an $XY$ antiferromagnet via its magnetic moment --- is dressed by the thermal fluctuations of the staggered magnetization. To the best of our knowledge, this is also the first demonstration of Bloch-Nordsieck-like physics with Goldstone bosons.



\begin{acknowledgments}
We thank Aydin Cem Keser and Yaroslav Kharkov for useful discussions. This work was supported by the Australian Research Council Centre of Excellence in Future Low Energy Electronics Technologies (CE170100039).
\end{acknowledgments}

\appendix

\section{Excitations of easy-plane ferromagnet \label{app:ferro}}

As pointed out in the main text of this article, it is well known that the $O(3)$ symmetric Heisenberg \textit{ferro}magnet has excitations with a quadratic dispersion, while those in an antiferromagnet have a linear dispersion. In this appendix, we demonstrate using spin wave theory that an easy-plane $O(2)$ symmetric ferromagnet also has a linear excitation spectrum.

Consider the Hamiltonian \eqref{eq:xyhamiltonian}, but now with $J = - \lvert J \rvert < 0$. The Heisenberg equations of motion for the $\hat{\mathbf{y}}$ and $\hat{\mathbf{z}}$ spin components are
\begin{align}
    \frac{\dd}{\dd t} S_n^{(y)} &= -J \sum_{\langle j,n \rangle} S_j^{(x)} S_n^{(z)}, \\
    \frac{\dd}{\dd t} S_n^{(z)} &= -J \sum_{\langle j,n \rangle} S_j^{(y)} S_n^{(x)} - S_j^{(x)} S_n^{(y)}.
\end{align}
If we choose the $\hat{\mathbf{x}}$ axis to coincide with the direction of spontaneous magnetization, then to leading order in $1/S$, we can set $S_m^{(x)} = S$ everywhere, so that
\begin{align}
    \frac{\dd}{\dd t} S_n^{(y)} &\simeq -4 J S S_n^{(z)}, \\
    \frac{\dd}{\dd t} S_n^{(z)} &\simeq -J S \sum_{\langle j,n \rangle} S_j^{(y)} - S_n^{(y)}.
\end{align}
Switching to Fourier space by making the replacement $S_n^{(\alpha)} \rightarrow S_{\mathbf{k}}^{(\alpha)} e^{i \mathbf{k}\cdot\mathbf{r}_n}$ yields
\begin{align}
    \frac{\dd}{\dd t} S_{\mathbf{k}}^{(y)} &\simeq -4 J S S_{\mathbf{k}}^{(z)}, \label{eq:app:sys1} \\
    \frac{\dd}{\dd t} S_{\mathbf{k}}^{(z)} &\simeq 4 J S (1 - \gamma_{\mathbf{k}}) S_{\mathbf{k}}^{(y)}, \label{eq:app:sys2}
\end{align}
where
\begin{equation}
    \gamma_{\mathbf{k}} = \frac{1}{2} \Big[\cos(k_x) + \cos(k_y) \Big],
\end{equation}
in units of the lattice spacing $a = 1$. The system of equations \eqref{eq:app:sys1} -- \eqref{eq:app:sys2} is trivial to solve, and has a characteristic frequency
\begin{equation}
    \omega_{\mathbf{k}} = 4 \lvert J\rvert S \sqrt{1 - \gamma_{\mathbf{k}}}.
\end{equation}
Finally, in the long wavelength limit, we see that the dispersion becomes linear: $\omega_{\mathbf{k}} \simeq c k$, with $c = 2 \lvert J \rvert S$.


\section{Proof of probability factorization \label{app:n_particles}}

In this appendix, we give a mathematical proof of the arguments leading to \eqref{eq:pn_tot}; the probability for a process involving $N$ quasiparticles factorizes into the two particle probability and $N - 2$ independent ``soft'' particles. We will only consider the process involving the excitation of $N$ particles by the source (a Stokes process with $\Delta > 0$), as the argument is completely general. The integrated probability is
\begin{align}
    P_N^{(S)}(\mathbf{k},\omega) &= \mathcal{S} \int \prod_{j=1}^{N} \frac{\dd^2 \mathbf{k}_j}{(2 \pi)^2} \frac{1 + n(\omega_{\mathbf{k}_j})}{2 \omega_{\mathbf{k}_j} \rho} \nonumber \\
    & \times (2 \pi)^3 \delta(\omega - \Sigma_j \omega_{\mathbf{k}_j}) \delta^{(2)}(\mathbf{k} - \Sigma_j \mathbf{k}_j), \label{eq:app:pn}
\end{align}
where $\mathcal{S} = 1/N!$ if we do not distinguish between particles. We eliminate the momentum $\delta$ function by integrating over $\mathbf{k}_N$, which leaves an energy $\delta$ function with the argument
\begin{equation}
    \omega - \sum_{j = 1}^{N - 1} \omega_{\mathbf{k}_j} - \Big\lvert \mathbf{k} - \sum_{j=1}^{N-1} \mathbf{k}_j \Big\rvert. \label{eq:app:delta_arg}
\end{equation}
We have already shown in the main text that one particle will carry most of the energy from the probe, and we will suppose that this particle will be the one labeled $N$, so that we may approximate \eqref{eq:app:delta_arg} as 
\begin{equation}
    \Delta - \sum_{j=1}^{N-1} (1 - \cos\theta_{\mathbf{k}_j}) k_j, \label{eq:app:delta_arg2}
\end{equation}
where $\theta_{\mathbf{k}_j}$ is the angle between $\mathbf{k}_j$ and the incoming momentum $\mathbf{k}$. Additionally, we have under the integrand in \eqref{eq:app:pn}
\begin{equation}
    \frac{1 + n(\omega_{\mathbf{k}_N})}{2 \omega_{\mathbf{k}_N} \rho} \simeq \frac{T}{2 \omega_{\mathbf{k}}^2 \rho},
\end{equation}
which we can factorize out. This explicitly distinguishes between one particle and the rest, so that at this point we have $\mathcal{S} = 1/(N-1)!$. For the case $N = 3$, the remaining integral is straightforward to evaluate exactly, and reduces to \eqref{eq:finite_T_3_particle} in the limit $\mu \ll \Delta$. For larger $N$, it is more convenient to further distinguish between the particle with energy $\sim \Delta$ and the rest. Supposing that this will be the one labeled $N-1$, we may further approximate \eqref{eq:app:delta_arg2} by
\begin{equation}
    \Delta - (1 - \cos\theta_{\mathbf{k}_{N-1}}) k_{N-1},
\end{equation}
and now $\mathcal{S} = 1/(N-2)!$. Therefore, we can eliminate the $\delta$ function by integrating over $k_{N-1}$, yielding
\begin{align}
    P_N^{(S)}(\mathbf{k},\omega) &\simeq \frac{1}{(N-2)!} \frac{T}{2 \omega_{\mathbf{k}}^2 \rho} \int \frac{\dd \theta_{\mathbf{k}_{N-1}}}{2 \pi} \frac{T}{2 \rho \Delta}  \nonumber \\
    &\qquad \times \prod_{j=1}^{N-2}\left( \int_0^\Delta \frac{\dd^2 \mathbf{k}_j}{(2 \pi)^2} \frac{1 + n(\omega_{\mathbf{k}_j})}{2 \omega_{\mathbf{k}_j} \rho} \right) \nonumber \\
    &= P_2(\mathbf{k},\omega) \times \frac{1}{(N-2)!} \left( \frac{T}{4 \pi \rho} \log \frac{\Delta }{\mu} \right)^{N-2},
\end{align}
precisely as in \eqref{eq:pn_tot}.

\bibliography{apssamp}

\end{document}